\documentclass[aps,superscriptaddress,pra]{revtex4-1}
\usepackage{amsmath,amssymb,graphicx,float,epstopdf}
\usepackage{dcolumn}
\usepackage{bm}
\usepackage{color}
\usepackage{setspace}

\DeclareMathOperator{\Bnabla}{\boldsymbol{\nabla}}
\DeclareMathOperator{\Bcdot}{\boldsymbol{\cdot}}
\DeclareMathOperator{\Bu}{{\bf u}}
\DeclareMathOperator{\Bn}{{\bf n}}
\DeclareMathOperator{\Br}{{\bf r}}

\begin{document}

\title{Control of droplet evaporation on smooth chemical patterns} 

\author{Michael Ewetola}
\affiliation{School of Mathematics and Statistics, The Open University, Milton Keynes MK7 6AA, United Kingdom}
\author{Rodrigo Ledesma-Aguilar}
\affiliation{Institute for Multiscale Thermofluids, School of Engineering, University of Edinburgh, The King's Buildings, Mayfield Road, Edinburgh EH9 3FB, United Kingdom}
\author{Marc Pradas}
\email[]{marc.pradas@open.ac.uk}
\affiliation{School of Mathematics and Statistics, The Open University, Milton Keynes MK7 6AA, United Kingdom}

\date{\today}

\begin{abstract}
We investigate the evaporation of a two-dimensional droplet on a solid surface. The solid is flat but with smooth chemical variations that lead to a space-dependent local contact angle. We perform a detailed bifurcation analysis of the equilibrium properties of the droplet as its size is changed, observing the emergence of a hierarchy of bifurcations that strongly depends on the particular underlying chemical pattern. Symmetric and periodic patterns lead to a sequence of pitchfork and saddle-node bifurcations that make  stable solutions to become saddle nodes. Under dynamic conditions, this change in stability suggests that any perturbation in the system can make the droplet to shift laterally while relaxing to the nearest stable point, as is confirmed by numerical computations of the Cahn-Hilliard and Navier-Stokes system of equations.  We also consider patterns with an amplitude gradient that creates a set of disconnected stable branches in the solution space, leading to a continuous change of the droplet's location upon evaporation.
\end{abstract}

\maketitle 

\section{Introduction}
The ability to control the configuration of a droplet evaporating on a solid surface is important for a wide range of applications, such as printing, coating, micro-patterning, and heat transfer~\cite{brutin2018recent}. 
One of the key issues is to understand how the properties of the solid affect the contact line of the droplet, i.e.~the line where all phases meet, and a substantial amount of work has been dedicated to that matter, see e.g.~\cite{picknett1977evaporation,Deegan_Nature_1997,hu2002evaporation,Ledesma_SM_2014,stauber_JFM_2014,dietrich2016stick,saenz2017dynamics,Amini_PRF_2017b,wray2020competitive}. 
In the ideal limit of a perfectly smooth and flat solid surface, a droplet keeps a constant shape characterised by the intersection angle of the liquid-gas interface with the solid. 
Such constant-contact-angle mode of evaporation implies the smooth retraction of the contact line as the droplet evaporates. 
In contrast, surfaces with microscopic defects,
either chemical or topographical, are able to induce the phenomenon known as contact-line pinning, whereby the translational motion of the 
contact line is suppressed~\cite{picknett1977evaporation}. 
Therefore, in the limiting situation of complete pinning, an evaporating droplet would exhibit a constant-contact-area mode of evaporation. 
In practice, a widely accepted view is that droplet evaporation proceeds either as a combination of these two limiting modes, often called a stick-slip mode of evaporation~\cite{stauber_JFM_2014} 
or as a combination of pinning and de-pinning of the contact line, called a stick-jump mode~\cite{dietrich2016stick}.

Recently, ultra-smooth smooth pinning-free surfaces which allow large-scale wettability patterns (comparable to the droplet size) have been developed. 
Such surfaces can be achieved by introducing an intermediary smooth layer that shields the droplet from the underlying solid surface, 
and include Slippery Liquid Infused Porous Surfaces (SLIPS)~\cite{Wong_Nature_2011,smith2013droplet,guan2017drop} and Slippery Covalently Attached Liquid Surfaces (SOCALS)~\cite{wang2016covalently}.
On flat SLIPS and SOCALS, a constant-contact-angle mode of evaporation has been reported, supporting the absence of contact-line pinning on these surfaces~\cite{guan2015evaporation,armstrong2019pinning}. 
However, introducing a large-scale topographical patterning has been shown to induce bifurcations between well-defined droplet configurations upon evaporation, 
which are  paced by dynamic “snap” events~\cite{Wells_NatCom_2018}.  
This has opened up the possibility to use solid surfaces with smooth wettability variations to control both the evaporation process and the motion of the droplet.

Here we study the evaporation of two-dimensional (2D) droplets on a perfectly flat and smooth, but chemically patterned surface.
We consider chemical patterns that lead to a smooth variation of the local equilibrium contact angle, thus eliminating pinning effects.  
The evaporation is assumed to be quasi-static and dictated by the equilibrium properties of the system, which depend on both the droplet's size and the 
%
specific chemical pattern of the substrate. 
By constructing the interfacial energy landscape of the system, we identify all possible equilibrium solutions of the droplet shape. 
On perfectly symmetric patterns, equilibrium solutions correspond to branches parametrised by the droplet's cross-sectional area, position and contact radius. 
Such branches form a network in the three-dimensional parameter space, where nodes correspond to pitchfork bifurcation points. 
Increasing the amplitude of the wettability pattern gives rise to folded nodes that signal the onset of saddle-node bifurcations.  
Introducing a weak bias in the pattern leads to a disconnection of the equilibrium branches and to the symmetry breaking of the pitchfork bifurcation nodes. 
%
Increasing the strength of the bias  creates a set of continuous branches of stable equilibrium solutions where the droplet's position varies smoothly upon changes in the cross-sectional area, 
suggesting that directed motion is possible on this type of surfaces. 

To understand the droplet dynamics upon evaporation on chemically patterned surfaces, we present numerical simulations of the Cahn-Hilliard and Navier-Stokes system of equations. 
We focus on the quasi-static regime, where droplet evaporation is dominated by diffusion into the gas phase.
For periodic and symmetric patterns, the droplet exhibits lateral movements when its cross-sectional area reaches the pitchfork bifurcations predicted by the theory, equivalent to the snap evaporation mode reported by Wells et al.~\cite{Wells_NatCom_2018}. 
On asymmetric patterns, 
the pitchfork branches are disconnected, and the droplet follows a smooth motion in a preferred direction as its size decreases in time, also in good agreement with the theory.  
%
%
Our results show that the interplay between a phase change and surface wettability can be exploited to control the motion of droplets on patterned solid surfaces in the absence of the anchoring effect of pinning.

\section{Equilibrium properties: Bifurcation analysis}
\label{s: Bifurcation}
\begin{figure}[t!]
\centering
\includegraphics[width=1.0\textwidth]{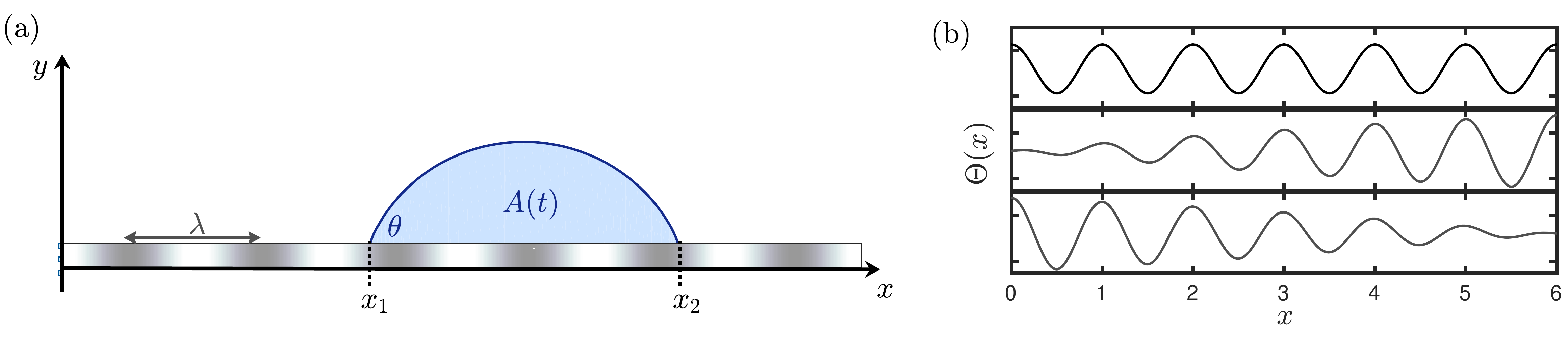}
\caption{(a) Two-dimensional droplet on a flat substrate with a smoothly varying chemical pattern. The location of the droplet's contact points is $x_1$ and $x_2$, and the contact angle is $\theta$. The droplet size $A(t)$ decreases in time and $\lambda$ is the wavelength of the periodic pattern. (b) Examples of chemical patterns, which are described by the local contact angle $\Theta(x)$.}	
\label{f: setup}
\end{figure}

Figure \ref{f: setup}(a) shows a {schematic representation of the system}
considered in this work. 
{A 2D droplet rests on a solid flat surface of non-uniform wettability.  
Here, we consider a periodic variation of the surface chemical properties along the lateral coordinate, $x$, which we model} using a spatially-dependent function, $\Theta(x)$, given by:
\begin{equation}
\cos \Theta (x)=\cos \theta_{0} - \epsilon \mathcal{F}(x), 
\label{e: Theta}
\end{equation}
where  $\theta_{0}$ is the reference homogeneous contact angle, $\epsilon$ controls the strength of the chemical pattern and $\mathcal{F}(x)$ is a generic periodic function. (We note that the reason to write $\cos\Theta(x)$ in the above equation is to simplify the analytical treatment presented below, see Eq.~(\ref{e: Energy general})). 

At equilibrium, and in the absence of gravity, the contact angle $\theta$ on both contact points of the droplet, 
$x_1$ and $x_2$, is the same and equal to the contact angle imposed by the chemical pattern, i.e.~$\theta= \Theta(\ell \pm R)$. 
{Here,} $\ell = (x_1+x_2)/2$ is the droplet shift and corresponds to the location of the droplet's midpoint 
{relative} to the origin $x=0$, and $R = (x_2-x_1)/2$ is the droplet footprint, see Fig.~\ref{f: setup}. 
Therefore, the shape of the free surface of the droplet, which we denote as $h(x)$, is given by a circular arc {whose cross-sectional area $A$} satisfies the relation:
\begin{subequations}
\label{e: Area}
\begin{align}
A&= \frac{R^{2}}{2} \frac{2\theta-\sin(2\theta)}{\sin^{2}\theta}, \\
\cos\theta & = \cos \theta_{0}-\epsilon\mathcal{F}(\ell\pm R). 
\end{align}
\end{subequations}
%
For a fixed droplet area, the stability of the equilibrium solutions  for $(\ell,R)$  that satisfy Eqs.~\eqref{e: Area} can be determined from the interfacial energy {(per unit length of the contact line)}
%
\begin{equation}
E(\ell,R)=\frac{2\gamma\theta R}{\sin \theta}-\gamma\int_{\ell -R}^{\ell +R} \cos\Theta(x) dx,
\label{e: Energy general}
\end{equation}
where $\gamma$ is the liquid/gas surface tension. Inserting Eq.~(\ref{e: Theta}) into Eq.~(\ref{e: Energy general}) gives:
\begin{equation}
E(\ell,R)=2\gamma R\left(\frac{\theta}{\sin \theta}-\cos \theta_{0}\right)  +\gamma\int_{\ell -R}^{\ell +R}\epsilon\mathcal{F}(x)\, dx,
\label{e: Energy}
\end{equation}
where $R$ and $\theta$ are given by Eqs.~(\ref{e: Area}). For a given droplet's area $A$,  we can compute the interfacial energy and find its extrema, which correspond to the equilibrium states of the droplet. 
In the following, we will analyse how the stability of the equilibrium states changes with the droplet area, leading to a hierarchy of bifurcation diagrams that are dictated by the underlying chemical pattern. 
These bifurcation diagrams will, in turn, inform about the possible (stable) trajectories in the $(A,\ell,R)$ space, which can be observed as the droplet's size is dynamically changed, see Sec.~\ref{s: Numerics}.

\subsection{Periodic and symmetric chemical patterns}
\label{ss: symmetric}
We start by considering periodic and symmetric patterns. 
{For simplicity, we consider} the the function $\mathcal{F}(x) = \cos(kx)$, where $k=2\pi/\lambda$ 
{and $\lambda$ is} the wavelength of the chemical variation.
We non-dimensionalise the system of equations (\ref{e: Area}) and (\ref{e: Energy}) by taking $\lambda$ as the typical length scale, such that the new dimensionless variables are $x' = x/\lambda$, $R'=R/\lambda$, $A' = A/\lambda^2$, and $E' = E/(\gamma\lambda)$. 
For convenience, we will drop the primes in the notation used in the rest of the paper. 
Under these conditions, Eq.~(\ref{e: Energy}) becomes:
\begin{equation}
E(\ell,R)=2R\left(\frac{\theta}{\sin \theta}-\cos \theta_{0}\right)  +\frac{\epsilon}{\pi} \sin(2\pi R)\cos(2\pi\ell),
\label{e: Energy periodic sym}
\end{equation}
where $\theta$ and $R$ are related through Eqs.~(\ref{e: Area}). 

We will now show that depending {on the strength of variation of the chemical pattern, given by the amplitude $\epsilon$},  different bifurcation points emerge as the droplet's size is changed.
{We anticipate two regimes: a small-$\epsilon$ regime, $\epsilon < \epsilon_c$, with $\epsilon_c$ a critical amplitude corresponding to a cusp point, below which all bifurcation points correspond to pitchfork bifurcations, 
and a large-$\epsilon$ regime, $\epsilon \geq \epsilon_c$, where a hierarchy of pitchfork and saddle node bifurcation develops.
} 

\subsubsection{Pitchfork bifurcation}
\begin{figure}[t!]
\centering
\includegraphics[width=1.0\textwidth]{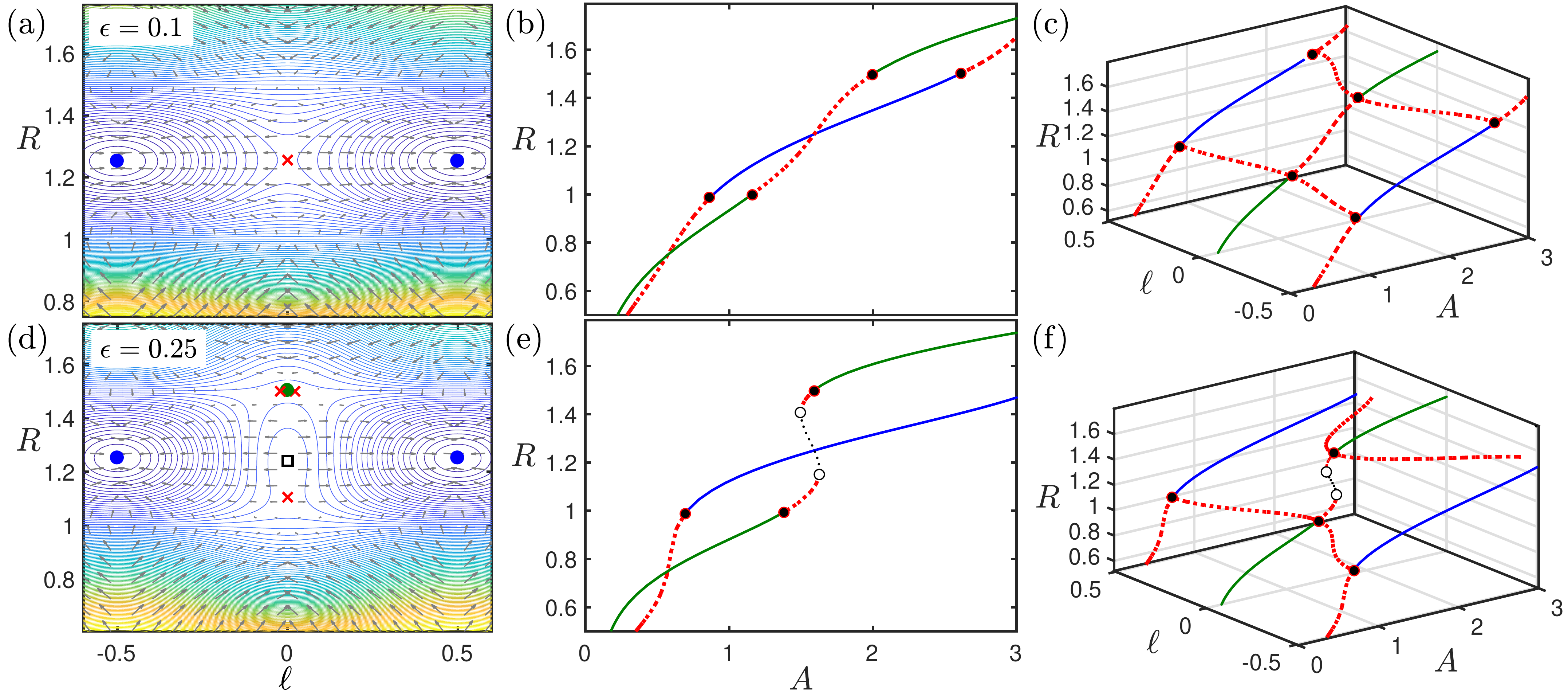}
\caption{Top panels (a,b,c) correspond to $\epsilon=0.1$ and lower panels (d,e,f) to $\epsilon=0.25$. (a,d) Interfacial energy contour plots for $A=1.5$, where energy levels increase from blue to yellow. Blue and green circles correspond to stable equilibrium solutions, crosses to saddle nodes that are stable to axisymmetric perturbations but unstable against lateral displacements, and empty squares correspond to unstable solutions. (b,e) Droplet lateral radius $R$ as function of the area $A$, where green and blue branches correspond to droplet stable solutions that are aligned with a maximum and a minimum  of the chemical pattern, respectively. The red dashed branches are saddle nodes and the black dotted branch corresponds to unstable solutions. (c,f) Bifurcation diagrams showing all possible solutions. Solid points represent subcritical pitchfork bifurcations and empty circles mark the onset of saddle-node bifurcations.}	
\label{f: pitchfork}
\end{figure}

We first consider the regime $\epsilon < \epsilon_c$.
We note that the work done in Ref.~\cite{pradas2016dynamics} analysed this case with a chemical pattern given by $\Theta(x) = \theta_0+\epsilon\cos(kx)$, reporting the emergence of subcritical pitchfork bifurcations on the ($\ell$, $A$) diagram. 
In this section, we revisit this case with the 
chemical pattern given by Eq.~(\ref{e: Area}b), 
which has the advantage that it leads to the explicit expression of the energy, Eq.~(\ref{e: Energy periodic sym}). 

Figure \ref{f: pitchfork}(a) shows a contour plot of the energy for a fixed 
droplet size, $A = 1.5$, and for $\epsilon=0.1$. 
%
{We first focus on equilibrium solutions for the droplet shape aligned with minima and maxima of the chemical pattern, marked with blue circles and red crosses in the figure.}
Solutions that are aligned with a minimum of the chemical pattern (i.e., $\ell = \pm (2n+1)/2$ for $n=0,1,2,\dots$) are stable, whereas solutions that are aligned with a maximum  ($\ell = \pm n$ for $n=0,1,2,\dots$) are saddle nodes, which are  stable to axisymmetric perturbations but unstable against lateral displacements along the solid surface. 
Therefore, if the droplet is on a saddle node, any perturbation on the system 
{will} destabilise the droplet's location and make the droplet shift laterally to either of the two stable solutions that are located to the left or right~\cite{pradas2016dynamics}. 

By fixing the location of the droplet to be  aligned with either a maximum or minimum of the chemical pattern, and by changing the droplet size $A$, we construct  two branches of solutions that are parametrised by the droplet's lateral radius $R$, as shown in Fig.~\ref{f: pitchfork}(b), where dashed lines correspond to saddle nodes and solid lines correspond to stable solutions. 
%
{The} stability of these solutions changes from stable to saddle node (or viceversa) at specific values of $A$. 
%
{Extending} this analysis to include droplet solutions that are located between minima and maxima of the chemical pattern, {yields} the three-dimensional bifurcation diagram shown in Fig.~\ref{f: pitchfork}(c). 
%
Stability transitions {correspond to} pitchfork bifurcations: a stable point (green solid line) collides with two saddle nodes to become a saddle node (subcritical pitchfork bifurcation), and a saddle node collides with two saddle nodes to become a stable solution (inverted subcritical pitchfork bifurcation). 
Therefore, in a dynamic {situation}, where the droplet's area is slowly decreasing in time, it is expected that around these bifurcation points, any perturbation that can break the plane symmetry will make the droplet shift and change location: if it is aligned with a maximum of the chemical pattern it will move to a minimum and vice versa. 

The critical droplet footprint $R_\mathrm{p}$ at which the pitchfork bifurcations occur can be determined explicitly by noting that at these points the stability of the solution changes from a stable to a saddle node.
Hence, these points satisfy $\partial_\ell^2E(\ell,R_\mathrm{p}) = 0$. Imposing this condition to Eq.~(\ref{e: Energy periodic sym}) {gives the relation} 
\begin{equation}
{\sin(2\pi R_\mathrm{p})=0,}     
\end{equation}
and hence the pitchfork critical radii are
\begin{equation}
R_\mathrm{p}=\frac{n}{2}, 
\label{e: Pitchfork radii}
\end{equation}	
for $n=1,2,\dots$. 
%
Therefore, pitchfork bifurcations occur at precise locations of the droplet's edges: either at minima or maxima of the chemical pattern. 
Remarkably, this geometrical property holds regardless of the chemical pattern, i.e., $R_p$ is independent of the homogeneous contact angle, $\theta_0$, and the amplitude of the substrate's chemical variation, $\epsilon$.
Instead, the effect of these parameters is to determine the critical contact angle $\theta_\mathrm{p}$ and area $A_\mathrm{p}$ at the bifurcation points, which follow from Eq.~(\ref{e: Area}).

\subsubsection{Cusp and saddle-node bifurcations}

We now study the regime $\epsilon\geq \epsilon_c$.
We observe multiple solutions for the same droplet area and midpoint location [see green circle, empty box and red cross at $\ell=0$ in Fig.~\ref{f: pitchfork}(d)]. 
Such solutions lie within S-shaped branches of the $R(A)$ curve characterised by two turning points [see Fig.~\ref{f: pitchfork}(e)]. 
These turning points mark the onset of saddle-node bifurcations whereby a saddle node solution collides with an unstable solution. 
{Such transitions are identified as empty circles in the three-dimensional bifurcation diagram shown in Fig.~\ref{f: pitchfork}(f).}

\begin{figure}[t!]
\centering
\includegraphics[width=1.0\textwidth]{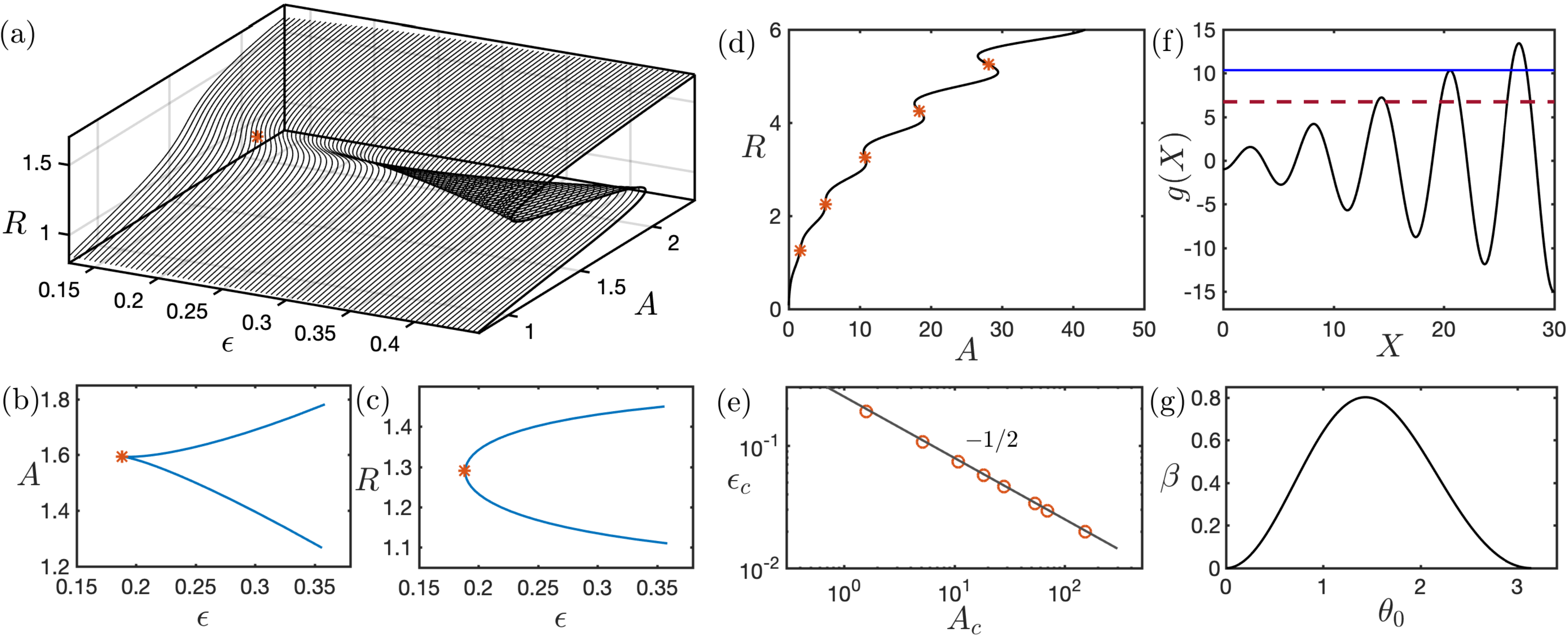}
\caption{(a,b,c) Emergence of a cusp bifurcation as the strength of the chemical pattern, $\epsilon$, is increased for the case with $\theta_0 =70^\circ$. Panel (a) shows the three-dimensional plot $(\epsilon, A, R)$, where the red asterisk marks the critical cusp point. Panels (b) and (c) show the corresponding projections onto the $(\epsilon, A)$ and $(\epsilon, R)$ planes, respectively. (d) Droplet footprint as function of its size for  $\epsilon=0.1$ and $\ell=0$. Red asterisk mark the critical radii at which a cusp bifurcation occurs. (e) Critical values of the strength of the chemical pattern to induce a cusp bifurcation as function of the droplet size. Solid line corresponds to a power law with exponent $-1/2$. (f) Plot of the function $g(X)$ where solid and dashed lines correspond to $\beta/\epsilon$ with $\theta_0=70^\circ$ and $\epsilon=0.046$ and $\epsilon = 0.058$, respectively. (g) Plot of the function $\beta(\theta_0)$. }	
\label{f: cusp}
\end{figure}

The emergence of unstable solutions is a consequence of a cusp bifurcation that occurs as $\epsilon$ is continuously increased, as is shown in Fig.~\ref{f: cusp}(a). 
At the critical cusp point, $\epsilon_c$, two new branches of solutions emerge, which correspond to the two turning points. 
Figs.~\ref{f: cusp}(b,c) show the evolution of these turning points on the $(\epsilon, A)$ and $(\epsilon, R)$ planes, in agreement with the standard form of the cusp bifurcation~\cite{Strogatz_1985}. 

It is important to note that the saddle-node bifurcations (i.e.~the turning points on the $R(A)$ curve) are not only observed as {$\epsilon$ increases, but also as the droplet size $A$ increases for a fixed value of $\epsilon$}, as is shown in Fig.~\ref{f: cusp}(d). 
We can understand this set of folds as a result of  a series of cusp bifurcations that occur at different critical points in the $(\epsilon, A,R)$ space, i.e.~each cusp bifurcation is described in terms of a critical strength $\epsilon_c$, a critical size $A_c$ and critical radius $R_c$. 

Fig.~\ref{f: cusp}(e) shows the set of critical amplitudes $\epsilon_c$ as a function of the critical droplet size $A_c$. 
To understand the scaling relation between $\epsilon_c$ and $A_c$, 
we note that for a fixed $\ell$, the turning points in Fig.~\ref{f: cusp}(d) are given by the stationary points of the 
function $A(R)$, i.e.~$dA/dR =0$, where $A(R)$ is given by Eq.~(\ref{e: Area}). In the limit of $\epsilon\to 0$ 
we write the expansion 
\begin{equation}
A = A_0(R)\left[1 +\frac{\epsilon}{\beta} \cos(2\pi R)\right]+{\rm h.o.t},
\label{A: exp}
\end{equation}
where $A_0=R^{2}(2\theta_0-\sin(2\theta_0)/2\sin^{2}\theta_0$ is the droplet size when $\epsilon = 0$, and we have defined the parameter 
\begin{equation}
\beta =\frac{ (\theta_0-\sin\theta_0\cos\theta_0)\sin\theta_0}{2(1-\theta_0\cot\theta_0)}.
\end{equation}
Imposing $dA/dR=0$ in Eq.~(\ref{A: exp}) and rearranging we find that the radii $R_s$ at which the saddle-node bifurcations occur are solutions of the equation
%
\begin{equation}
\label{e: Asympt R Cusp}
\frac{X_s}{2}\sin X_s  -\cos X_s= \frac{\beta (\theta_0)}{\epsilon},
\end{equation}
where $X_s = 2\pi R_s$. 
 In addition, we note that the stationary points $X_c$ of the function $g(X) = (X/2)\sin X-\cos X$ correspond to the onset of the cusp bifurcation. 
 Therefore, the critical value $\epsilon_c$ where the cusp bifurcation occurs, can be obtained by imposing the condition $g(X_c) = \beta/\epsilon_c$.  
 
 Figure \ref{f: cusp}(f) shows a plot of the function $g(X) = (X/2)\sin X-\cos X$, and the constant $\beta/\epsilon$ for $\theta_0=70^\circ$ and two arbitrary values of $\epsilon$. 
 For $\epsilon=0.046$ the plot shows a cusp bifurcation that corresponds to the first intersection between the constant $\beta/\epsilon$ (blue solid line) and the function $g(X)$ at the maximum $X_c\approx 20$. 
 Increasing the amplitude of the chemical pattern to $\epsilon=0.058$ around $X=20$, the constant $\beta/\epsilon$ (red dashed line) intersects $g(X)$ at two points that correspond to the saddle node bifurcations. 

Expanding the function $g(X)$ around $X_c$, we find that the solutions near the cusp bifurcation are given by
\begin{equation}
X \simeq X_c\pm \delta\left(\frac{1}{\epsilon}-\frac{1}{\epsilon_c}\right)^{1/2},
\end{equation}
where $\delta = \sqrt{2\beta(\theta_0)/g''(X_c)}$ is a constant that depends on $\theta_0$ only, and the critical value is given by
\begin{equation}
\epsilon_c = \frac{\beta(\theta_0)}{g(X_c)}\sim \beta(\theta_0) A_c^{-1/2},
\end{equation}
where we have approximated $g(X_c)\sim X_c/2$, transformed back to the radius variable $R$, and made use of the fact that  at the onset of the cusp bifurcation $A\sim R^2$. The above relation is in agreement with the scaling behaviour shown in Fig.~\ref{f: cusp}(e). 
%
Because $\beta$ is always finite [cf.~\ref{f: cusp}(g)], an important conclusion is that cusp, and, consequently saddle node bifurcations are observed for any wetting condition, as long as $\epsilon\ne 0$.
In addition, because the critical cusp area $A_c$ is normalised by the squared wavelength $\lambda^2$, we conclude that, for a fixed droplet area, cusp bifurcations are favoured in the microscopic limit of $\lambda\to 0$.  
 
\subsection{Patterns with an amplitude gradient}
\label{ss: asymmetric theory}

The results shown in Fig.~\ref{f: pitchfork} indicate that, {on symmetric chemical patterns, a droplet will adopt equilibrium configurations which are aligned with either a maximum or a minimum of the pattern.}  
As the droplet's size {changes,} 
the stability of {such}
configurations alternates between stable and saddle nodes through a sequence of pitchfork bifurcations that can promote droplet lateral motion: any perturbation that breaks the plane symmetry will make the droplet change from a saddle node to a stable location {where the interfacial energy is at a minimum}.  
However, and because of symmetry, there is no bias for the change in position of the droplet,
hence ruling out the possibility to induce droplet motion towards a \emph{preferred} direction. 
To this end, here we explore a non-symmetrical chemical pattern with the aim to 
{determine whether it is possible to achieve}
directed displacement in the droplet's location as the droplet size is changed.

We consider a pattern with an amplitude gradient described by the function:
\begin{equation}
\mathcal{F}(x) = \frac{2}{\pi}\arctan\left(\frac{x}{L}\right)\cos(2\pi x),
\label{e: gradient}
\end{equation}
where $L$ is the length over which the gradient varies. An example of the above pattern with $\epsilon = 0.2$ and $L=6$ is shown in Fig.~\ref{f: setup}(b). (We note that the change of sign of the gradient can be imposed by replacing $x$ by $L-x$ in the argument of the $\arctan$).

\begin{figure}[t!]
\centering
\includegraphics[width=1.0\textwidth]{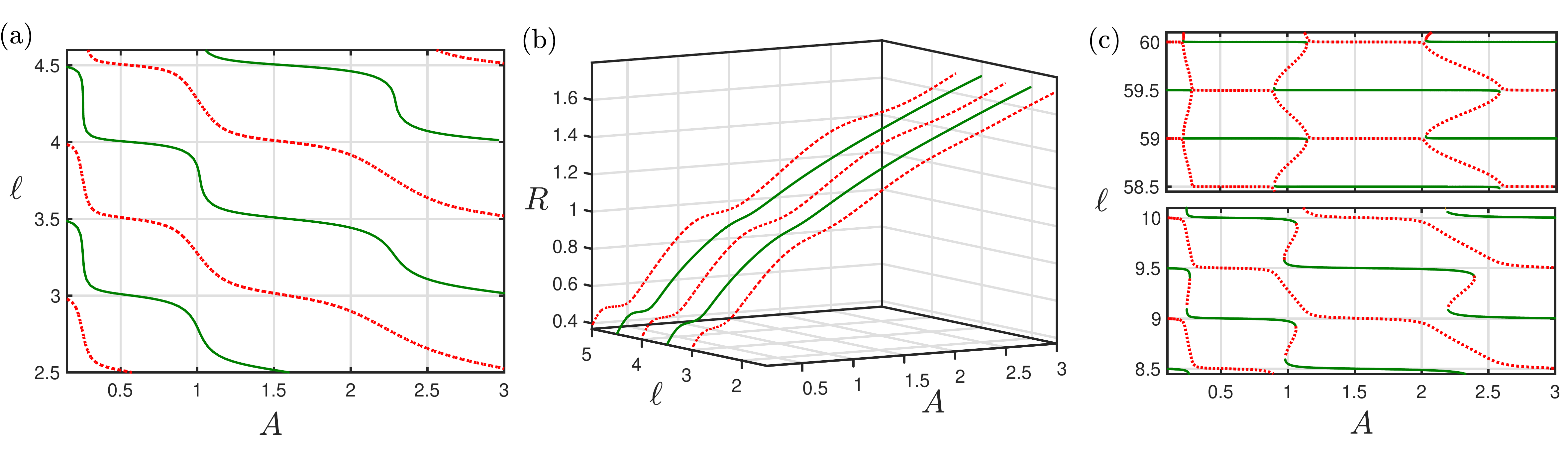}
\caption{Bifurcation diagrams for the case of a chemical pattern with an amplitude gradient, given by Eq.~(\ref{e: gradient}) with $\epsilon =0.1$ and $L=6$. Panel (a) shows the equilibrium solutions of the droplet's midpoint location, $\ell$, as function of its size $A$ and panel (b) the bifurcation diagram in the three-dimensional space $(A,\ell,R)$. Solid green lines correspond to stable solutions and dashed red lines correspond to saddle nodes. Panel (c) shows the bifurcation diagram on the $(A,\ell)$ plane when $\ell>L$ (bottom) and $\ell\gg L$ (top).}	
\label{f: gradient}
\end{figure}

We solve Eqs.~(\ref{e: Area}) alongside Eq.~(\ref{e: gradient}) to find the equilibrium solutions for a given droplet size. Following the same procedure as in the previous section, we  construct the bifurcation diagrams as the droplet size is changed. Figure \ref{f: gradient} shows the branches of solutions on the $(A,\ell)$ plane and in the $(A,\ell,R)$ space. 
We observe that the lack of symmetry of the chemical pattern leads to a topological change in the bifurcation diagrams, characterised by a series of disconnected branches, which are either stable or saddle node. In particular, we can see that as the droplet size is changed, there always exists a set of stable branches that can be continuously parametrised by the droplet's midpoint, i.e.~$\ell(A)$.  
%
This implies that changing
the droplet size 
{can }lead to {a continuous} lateral displacement {along} a preferred direction. 

To understand {the onset of symmetry breaking and the consequent} topological change in the bifurcation diagram, 
{let us focus on} 
the case of $\ell>L$, noting that in the limit of $\ell\gg L$, the chemical pattern given by Eq.~(\ref{e: gradient}) becomes symmetric and equivalent to the case considered in the previous section. The bottom panel of Fig.~\ref{f: gradient}(c) shows the emergence of turning points along the stable branches for $\ell>L$, which in the limit of $\ell\gg L$ (top panel),  become pitchfork bifurcation points, thereby connecting the two previously disconnected stable and saddle node branches. This shows how the topological change in the bifurcation diagrams is purely controlled by the  degree of asymmetry of the chemical pattern. 

\section{Droplet evaporation}

In this section we study the  evaporation of a 2D droplet  on a solid surface. 
We assume that evaporation is quasi-static and driven by mass diffusion in the gas phase; hence, we neglect the effect of a temperature difference between the solid, liquid and gas phases.  
To model such a system, we adopt a diffuse-interface formulation that includes a wetting boundary condition at the solid substrate as well as an open boundary to drive the evaporation of the droplet.

\subsection{Diffuse-Interface formulation}
\label{s: DI}
We consider the Cahn-Hilliard and Navier-Stokes (CH-NS)  system of equations for an incompressible fluid:
\begin{subequations}
\label{e: CH-NS Dim}
\begin{align}
\frac{\partial \phi}{\partial t} + \Bu\Bcdot\Bnabla\phi  & = M\nabla^2\eta(\phi), \label{e:CH1} \\
\rho\left(\frac{\partial \Bu}{\partial t} +(\Bu\Bcdot\Bnabla)\Bu\right) &= -\Bnabla p +\mu\nabla^2\Bu -\phi \Bnabla \eta,  \label{e:NS1} \\
\Bnabla\cdot\Bu & = 0,
\end{align}
\end{subequations}
where $\Bu$ is the velocity field, $p$ is the pressure, $\rho$ the density, $\mu$ the dynamic viscosity, and $M$ the mobility parameter. 
%
The above equations are integrated in a two-dimensional domain $\Omega$ with boundary $\partial\Omega$, where $\phi$ is a locally conserved field that plays the role of an order parameter by taking two equilibrium limiting values, $\phi=+\phi_\mathrm{e}$ and $\phi=-\phi_\mathrm{e}$, which represent the liquid and vapour phases, respectively. 
Hence, in the following we identify the location of the interface as the level curve $\phi=0$. 

We define the chemical potential field, $\eta=\delta \mathcal{F}[\phi]/\delta \phi$, where $\mathcal F$ is the free-energy of the system:
\begin{equation}
\label{e:Free Energy}
\mathcal{F}[\phi] = \int_\Omega\frac{\sigma}{\xi}\left( F_b(\phi) +\frac{\xi^2}{2}\vert\Bnabla\!\phi\vert^2\right)\, d\Omega+\int_{\partial\Omega} F_w(\phi)\,ds.
\end{equation}
{Here,} {$F_b(\phi) = (1 -\phi^{2})^{2}/4$ is a double-well potential and $F_w(\phi)$ is the wall component of the free energy that models fluid/solid (wetting) interactions. The parameter $\sigma = (3/2\sqrt{2})\gamma$ is related to the surface tension $\gamma$, and $\xi$ is a small parameter controlling the width of the diffuse interface, such that in the limit of $\xi\to 0$, one recovers the macroscopic sharp interface formulation~\cite{Seppecher_IJES_1996, Anderson_ANRFM_1998,Yue_JFM_2010,Sibley_JFM_2013}.}
Minimization of the free energy (\ref{e:Free Energy}) gives:
\begin{equation}
\label{e:ChemPot}
\eta= \frac{\sigma}{\xi}\left(F'_{b}(\phi) -\xi^{2} \nabla^2 \phi\right),
\end{equation}
defined in $\Omega$ alongside the natural boundary condition:
\begin{equation}
\sigma\xi\left(\Bn\Bcdot\Bnabla\phi\right)=-F'_w,
\label{e:NaturalBoundary}
\end{equation}
which is applied on $\partial\Omega$.
It is convenient to non-dimensionalise Eqs.~(\ref{e: CH-NS Dim})-(\ref{e:NaturalBoundary}) by choosing the following dimensionless variables:
\begin{equation}
\Br^* = \frac{\Br}{L},\quad \Bu^* = \frac{\Bu}{U}, \quad t^* = \frac{Ut}{L},\quad p^* = \frac{p}{\rho U^2},\quad \eta^* = \frac{\eta}{\eta_0},\quad \phi^* = \frac{\phi}{\phi_\mathrm{e}},
\end{equation}
where $L$, $U$, and $\eta_0=\sigma/L$ are the typical length, velocity, and chemical potential scales of the system, obtaining
\begin{subequations}
\label{e: CH-NS}
\begin{align}
\frac{\partial \phi}{\partial t} + \Bu\Bcdot\Bnabla\phi  & = \frac{1}{Pe}\nabla^2\eta, \label{e:CH2} \\
\eta & = \frac{1}{Cn}\left(-\phi+\phi^{3} -Cn^2 \nabla^2 \phi\right), \label{e:mu}\\
\frac{\partial \Bu}{\partial t} +(\Bu\Bcdot\Bnabla)\Bu  &= -\Bnabla p + \frac{1}{Re}\nabla^2\Bu -\frac{1}{We} \phi\Bnabla \eta, \label{e:NS2}
\end{align}
\end{subequations}
alongside the continuity equation $\Bnabla\cdot\Bu = 0$. For simplicity, we have dropped the asterisks in the dimensionless variables and we have taken $\phi_e=1$. The set of dimensionless parameters in the above equations are defined as:
\begin{equation*}
Pe =\frac{UL^2}{M\sigma},\qquad Cn = \frac{\xi}{L},\qquad Re = \frac{\rho LU}{\mu},\qquad We = \frac{\rho U^2L}{\sigma},
\end{equation*} 
which correspond to the Peclet number, Cahn number, Reynolds number, and Weber number, respectively. Following the work reported in \cite{Ding_JCP_2007,Magaletti_JFM_2013}, the Peclet number is chosen to be inversely proportional to $Cn^2$, and throughout this study is set to $Pe =
1/3Cn^{2}$. For the other parameters, we take the values of $Re = 1 $, $We = 0.2$, and $Cn = 0.01$.  

In this formulation we choose $F_w$ to be a linear function in $\phi$~\cite{Cahn_JCP_1977,deGennes_RMP_1985,Aymard_JCP_2019},  given by $F_w(\phi) = -(\sqrt{2}\sigma/3)\cos\Theta(x)\phi$, where $\Theta(x)$ is the local equilibrium contact angle, which we assume that it may depend on the position $x$. After non-dimensionalisation, the boundary condition at the solid/fluid wall given by Eq.~\eqref{e:NaturalBoundary}, becomes
\begin{equation}
\Bn\Bcdot\Bnabla\phi=\frac{\sqrt{2}}{3Cn}\cos\Theta(x).
\label{e:Wetting}
\end{equation}
To drive slow evaporation and dynamically change the size of the droplet, we impose a fixed flux at the top of the system by imposing a Neumann's boundary condition for the chemical potential:
\begin{equation}
\Bn\cdot\Bnabla\eta\vert_{y=y_w} = -\eta_w,
\label{e: mu BC}
\end{equation}
where $y_w$ corresponds to the location of the top boundary, and $\eta_w>0$ is the imposed value for the chemical potential, noting that for $\eta_w = 0$ the system is closed. The system of equations and boundary equation conditions is solved by making use of finite elements (see Appendix).
\begin{figure}[t!]
\centering
\includegraphics[width=1.0\textwidth]{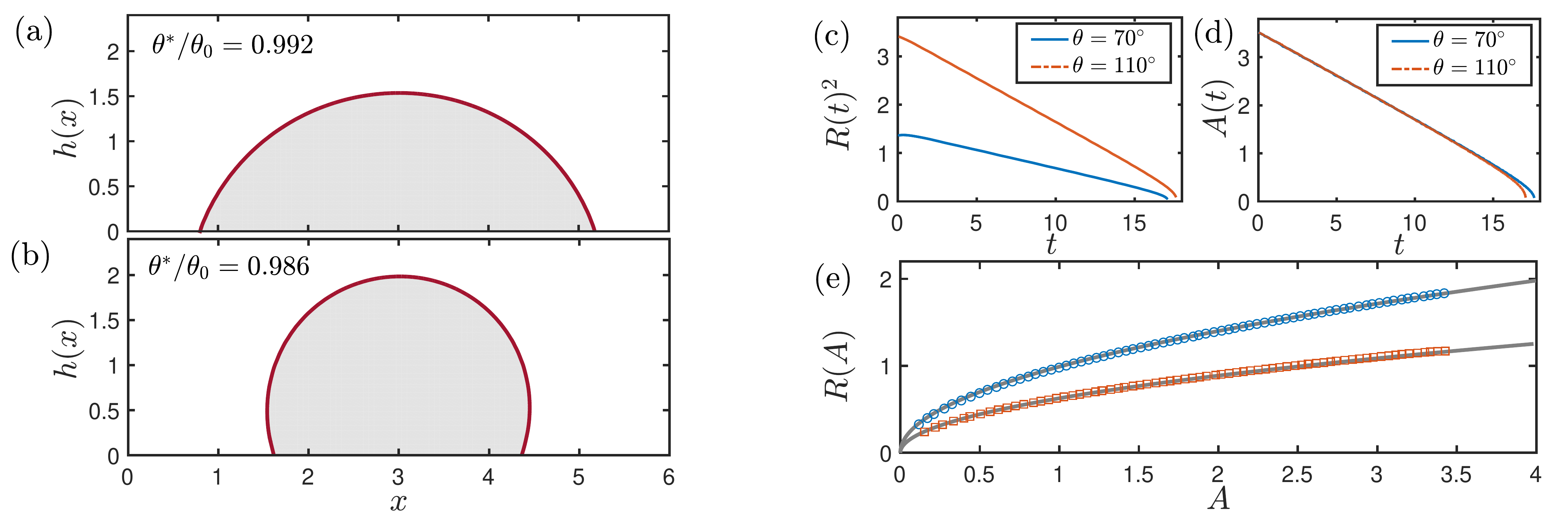}
\caption{(a,b) Equilibrium droplet shapes on homogeneous substrates with different wetting properties. The contact angle imposed by condition (\ref{e:Wetting}) is denoted as $\theta_0$ and the angle calculated numerically as $\theta^*$. (c,d) Time-dependent evolution of the squared droplet lateral radius, $R(t)^2$, and size, $A(t)$, for two different wetting properties. (e) Comparison between the numerically computed droplet lateral radius as function of its size and the theoretical expression given by Eq.~(\ref{e: Area}a) with $\theta = \theta_0$ (solid gray lines). Blue circles and red squares correspond to $\theta_0 = 70^\circ$ and $\theta_0 = 110^\circ$, respectively. }	
\label{f: numerics1}
\end{figure}

{To validate the numerical model, we first carry out simulations of the equilibrium state of droplets}
on {solid} substrates of {uniform} wetting properties. 
Figures \ref{f: numerics1}(a,b) show the equilibrium shapes for a hydrophilic and a hydrophobic homogeneous surface with $\Theta =70^\circ$ and $\Theta =110^\circ$, respectively. 
The contact angle was calculated numerically from the computations, and is in excellent agreement with that imposed by condition (\ref{e:Wetting}).  
We then impose the open flux boundary condition (\ref{e: mu BC}) with $\eta_w =4$ to drive evaporation. Figures \ref{f: numerics1}(c,d) show the time evolution of the squared lateral radius, $R(t)^2$,  and size $A(t)$,  and show that the droplet's footprint decreases in time as $R(t)\sim t^{1/2}$. 
{Fig. \ref{f: numerics1}(e) shows a parametric plot of the instantaneous radius vs cross-sectional area of the droplet. At all times, the simulation data follows the equilibrium geometrical relation given Eq.~(\ref{e: Area}a), hence confirming that the evaporation of the droplet is quasi-static.}
%

\subsection{Droplet evaporation on symmetric patterns}
\label{s: Numerics}
\begin{figure}[t!]
\centering
\includegraphics[width=1.0\textwidth]{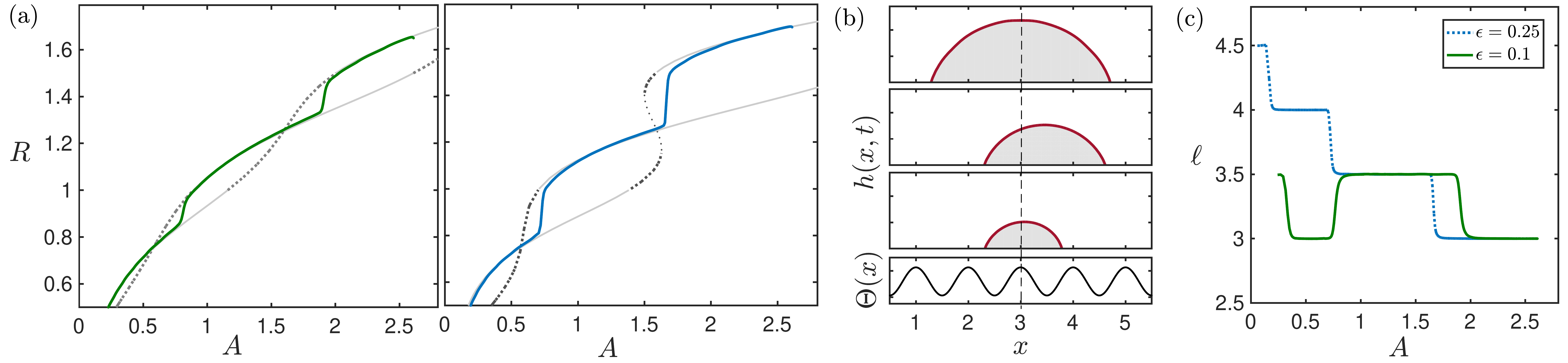}
\caption{Numerical simulations of slow evaporation by soling the CH-NS system of equations, Eqs.~(\ref{e: CH-NS}), with a symmetric chemical pattern, and with an evaporation rate of with $\eta_w = 2$. (a) Solid lines show the time evolution of the droplet footprint as function of the droplet size for $\epsilon=0.1$ (left) and $\epsilon = 0.25$ (right). The underlying gray lines correspond to the theoretical bifurcation diagrams shown in Fig.~\ref{f: pitchfork}. Panel (b) shows droplet snapshots at different times 
and panel (c) shows the evolution of the droplet's midpoint as function of the droplet size.}
\label{f: numerics - slow}
\end{figure}

We first consider a droplet evaporating on a symmetric chemical pattern given by Eq.~(\ref{e: Theta}) with $\mathcal{F}(x) = \cos(2\pi x)$ {and $\epsilon=0.1$.} 
The droplet is initially aligned with a maximum of the chemical pattern. 
%
We set the
evaporation rate to
$\eta_w=2$.
Figure \ref{f: numerics - slow}(a), left panel, shows that, as the droplet size decreases quasi-statically, the evolution of the lateral radius $R(A)$ is in excellent agreement with the trajectory predicted by the theoretical bifurcation diagram (shown in gray lines).

For droplet sizes larger than the critical value $A_\mathrm{p}$, which marks the onset of a pitchfork bifurcation, the droplet is fully stable and aligns with the maximum of the chemical pattern. 
When $A<A_\mathrm{p}$, the droplet solution becomes unstable against asymmetric perturbations and any small perturbation ({in the present case, numerical noise}) is able to break the plane symmetry forcing the droplet to shift laterally 
to a stable branch of solutions, which are aligned with a minimum of the chemical pattern and are located either to the left or right of the droplet's original location ($\ell=3$) [see Fig.~\ref{f: numerics - slow}(b)].   
The droplet then continues following the bifurcation diagram in this new location until another pitchfork bifurcation occurs, forcing the droplet to shift and to be aligned with a maximum again. 

The trajectory of droplet's midpoint as a function of its size is shown in Fig.~\ref{f: numerics - slow}(c), where we can see that lateral movements occur over a much faster time-scale than the timescale of evaporation. Such fast lateral movements correspond to the snap events that have been reported on topographical smooth surfaces~\cite{Wells_NatCom_2018}. 
We note that a similar behaviour is observed for larger values of the strength of the chemical pattern [see Figs.~\ref{f: numerics - slow}(a,c) for $\epsilon=0.25$ and Supplementary Movie 1]. 
It is important to remark that, because of the symmetry of the chemical pattern, the direction taken by the droplet at each pitchfork bifurcation is not predictable and hence cannot be controlled, i.e.~the droplet can shift either to the right or to the left.

\subsection{Droplet evaporation on asymmetric patterns}
\begin{figure}[t!]
\centering
\includegraphics[width=1.0\textwidth]{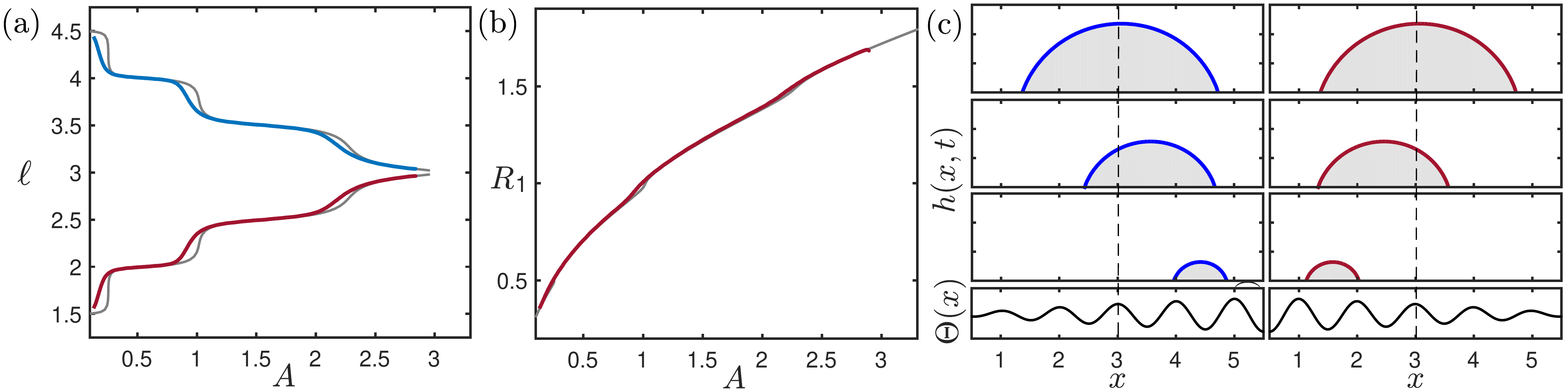}
\caption{Numerical simulations of the CH-NS system of equations, Eqs.~(\ref{e: CH-NS}), with a chemical pattern given by Eq.~(\ref{e: gradient}) with $\epsilon=0.1$ and $L=6$. Panel (a) shows the evolution of the droplet's midpoint as function of its size for a positive gradient (blue solid line) and negative gradient (red solid line). The underlying gray lines correspond to the stable solutions predicted by the theory. Panel (b) shows the droplet's footprint versus its size (red solid line) compared to the theoretical prediction (gray line). Panels in (c) show the corresponding  droplet profiles at different times.}	
\label{f: numerics - gradient}
\end{figure}
In this section we study the evaporation of a droplet on an asymmetric pattern.
We impose a chemical pattern with an amplitude gradient, described by  Eq.~(\ref{e: gradient}) where the amplitude of the chemical pattern gradually increases or decreases with $x$. 
Figure \ref{f: numerics - gradient}(a) shows the trajectories of the droplet's midpoint as the droplet size decreases for the case with a positive gradient (blue solid line) and negative  gradient (red solid line). We observe that in both cases, the asymmetry of the chemical pattern induces a continuous change in the droplet's midpoint location, forcing the droplet to move either to the left or right as its size decreases in time. 

We note that, as predicted by the theoretical analysis shown in Fig.~\ref{f: numerics - gradient}, the bifurcation diagrams for this type of chemical patterns consists  of a series of disconnected branches, which are either stable or saddle node. If the droplet is initially located at a stable location, it will remain on this branch during the entire process and continuously move following the stable branch of solutions, as it is observed in Fig.~\ref{f: numerics - gradient}(a). It is worth noting that in both cases of the gradient sign the droplet's footprint  decreases continuously in time following the same trajectory on the $(A,R)$ plane, as predicted by the theory [see Fig.~\ref{f: numerics - gradient}(b)].
Counter-intuitively, it is seen that the droplet moves towards higher amplitude of the chemical pattern, moving to the right with positive gradient and to the left with negative gradient, see Fig.~\ref{f: numerics - gradient}(c) and Supplementary Movies 2A and 2B. 
\begin{figure}[t!]
\centering
\includegraphics[width=1.0\textwidth]{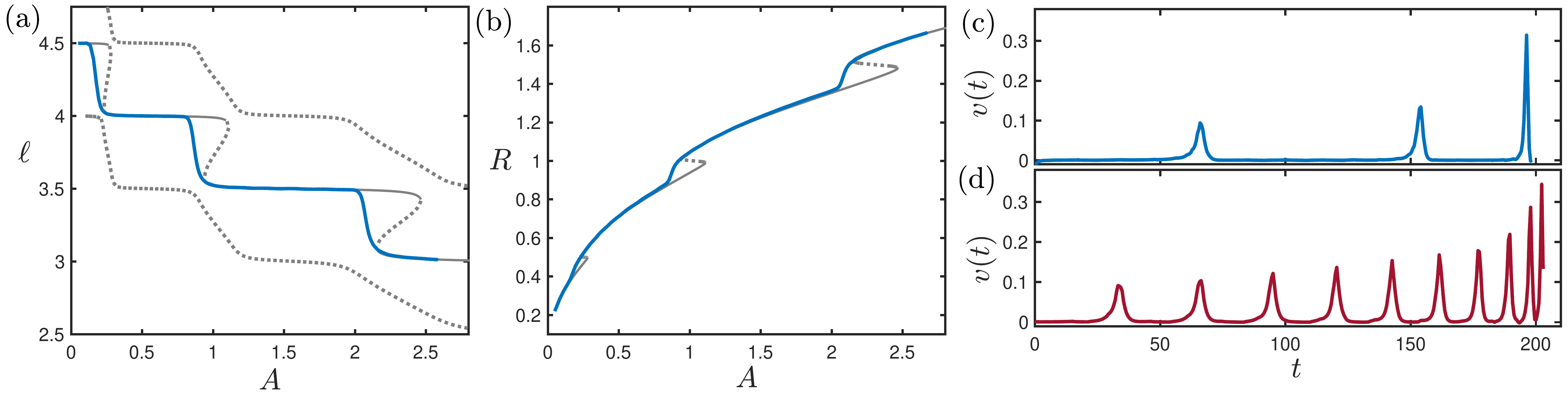}
\caption{Numerical simulations of the CH-NS system of equations, Eqs.~(\ref{e: CH-NS}), chemical pattern given by Eq.~(\ref{e: gradient}) with $\epsilon=0.1$ and $L=0.6$. Panel (a) shows the evolution of the droplet's midpoint as function of its size and the underlying gray lines correspond to the stable solutions predicted by the theory. Panel (b) shows the droplet's footprint versus its size (blue solid line) compared to the theoretical prediction (gray line). Panel (c) shows the speed $v(t)$ of the droplet's midpoint versus time (dashed blue line). Panel (d) shows the same speed but for a chemical pattern with smaller wavelength $\lambda$.}
\label{f: asymmetric - snap}
\end{figure}

By setting $L=0.6$ and keeping the same numerical domain size we approach the limit described in Sec.~\ref{ss: asymmetric theory}, Fig.~\ref{f: gradient}(c), in which the chemical pattern is nearly symmetric. 
Figures \ref{f: asymmetric - snap}(a,b) show the trajectories of the droplet's midpoint and footprint as function of the droplet size, respectively. We can recognise the presence of turning points on the $\ell(A)$ trajectory (see gray lines in Fig.~\ref{f: asymmetric - snap}(a)) leading to a rapid change in both the droplet's midpoint and footprint, similar to the snap events observed under symmetric patterns. However, because there is now a symmetry breaking the change in droplet's location is induced by imperfect pitchfork bifurcations. Hence, all movements are directed towards the same direction allowing for a better control of  droplet's position. 

The presence of snap events is clearly demonstrated in  Fig~\ref{f: asymmetric - snap}(c), where we plot the speed of the droplet's midpoint, $v(t) = \dot{\ell}(t)$. 
We can see that droplet's lateral movements become faster as the droplet size decreases. 
Decreasing the wavelength of the chemical pattern (but keeping the same initial droplet size, i.e., effectively increasing the dimensionless variable $A$) leads to a dynamics with a higher rate of lateral shifts, as expected (see Fig.~\ref{f: asymmetric - snap}(d) and Supplementary Movies 3 and 4).

\section{Concluding remarks}
We have presented analytical and computational results on quasi-static evaporation of a 2D droplet on a flat, chemically patterned surface. We considered patterns that are pinning free but have a smooth and periodic variation of the local equilibrium contact angle. We have shown that symmetric  patterns lead to a hierarchy of bifurcations in the three-dimensional parameter space represented by the droplet's cross sectional area, midpoint, and footprint. For an amplitude $\epsilon$ of the chemical pattern smaller than a critical value $\epsilon_c$ the nodes of the network correspond to pitchfork bifurcations that mark transitions between stable and saddle points. For $\epsilon>\epsilon_c$ a cusp bifurcation occurs leading to the emerge of turning points that mark the onset of saddle-node bifurcations. 

A detailed bifurcation analysis has revealed that pitchfork bifurcations occur at well defined locations of the chemical pattern, which are independent of the homogeneous contact angle and amplitude of the chemical variation. We have also shown that the amplitude critical value scales with the droplet's size as $\epsilon_c\sim A_c^{-1/2}$, hence suggesting that cusp bifurcations are favoured in the microscopic limit. Introducing a bias in the chemical pattern leads to a topological change in the bifurcation diagrams, whereby equilibrium solutions are characterised by disconnected branches in the parameter space. Such branches, which can be either stable or saddle points, are continuously parametrised by the droplet's midpoint, i.e.~$\ell(A)$, implying that changing the droplet's size may lead to a continuous lateral displacement. 
 
We have studied droplet dynamics upon evaporation by making use of the Cahn-Hilliard and Navier-Stokes system of equations. Periodic and symmetric patterns lead to a sequence of snap events where the droplet exhibits rapid lateral movements when its cross-sectional area reaches the pitchfork bifurcations predicted by the theory. This shows that {\it snap evaporation}~\cite{Wells_NatCom_2018} is also observed on planar surfaces with smooth chemical patterns.  In asymmetrical chemical patterns, the presence of disconnected branches leads to a smooth droplet's motion where its location continuously changes towards one direction, hence showing that droplet's motion can be controlled upon evaporation. In the limit of weak bias, the droplet dynamics is characterised by snap events but because of the slight symmetry breaking of the chemical pattern, they always occur towards the same direction. We have also shown that the maximum droplet's speed during a snap event increases as the droplet's size decreases.

The  ideas presented here can be used in applications of droplet control and mass transport. We have shown that well designed chemical patterns can lead to a well controlled motion  of the droplet as its size changes in time. It is important to note that our results have focused on evaporation process but the analysis presented here is equally applicable to other physical processes, such as condensation or mass transfer.

\begin{acknowledgments}
We acknowledge financial support by the UK Engineering and Physical Sciences Research Council (EPSRC) through Grant No.~EP/R041954/1.
\end{acknowledgments}

\appendix*

\section{Numerical method}
\label{ap: numerical method}

The system of the Cahn-Hilliard and Navier-Stokes  equations, Eqs.~(\ref{e:CH2} - \ref{e:NS2}),  is solved using finite elements. To this end, these equations are expressed in variational form (weak formulation) which is obtained by multiplying each equation by a test function and integrating the resulting equation over the domain $\Omega$.  The unknown functions to be approximated are referred to as  trial functions. 

To obtain the weak formulation of the system (\ref{e:CH2} - \ref{e:NS2}), we define test functions  $(\Phi,\Psi,\Xi,\Pi) \in \hat{W} \times \hat{W} \times \hat{W} \times \hat{W}$ corresponding to the trial functions  $(\phi,\mu,u,p) \in W \times W \times W \times W$, where $W$ and $\hat{W}$ are the trial and test spaces defined as:
\begin{align*}
&W = \{v\in H^{1}(\Omega) : v = u_{D} \ \ \ on\ \ \ \partial \Omega \},\\
&\hat{W} = \{v\in H^{1}(\Omega) : v = \hat{u_{D}}\ \ \ on\ \ \ \partial \Omega \},
\end{align*}
respectively, where, $\Omega$, $\partial \Omega$, $u_{D}$, $\hat{u_{D}}$ and $H^{1}(\Omega)$ are the spatial domain, boundary of $\Omega$, the value of the trial and test functions at $\partial \Omega$, and the  Sobolev space respectively. We first consider the Cahn-Hilliard equation by multiplying Eqs.~(\ref{e:CH2}) and  (\ref{e:mu}) by  $\Phi$ and $\Psi$, respectively, and integrating over the whole domain $\Omega$, to obtain:
\begin{align}
&  \langle \frac{\partial \phi}{\partial t} , \Phi \rangle+ \left< \Bu\Bcdot\Bnabla\phi , \Phi \right>  +  \frac{1}{Pe} \left\langle \Bnabla\eta, \Bnabla \Phi \right\rangle - \frac{1}{Pe} \left\langle \Bn \Bcdot \Bnabla \eta , \Phi \right\rangle_{\partial \Omega}=0, \label{e:CH3} \\
&\left\langle \eta, \Psi \right\rangle + \left\langle \left(\frac{\phi-\phi^{3}}{Cn}\right), \Psi \right\rangle -Cn \left\langle \Bnabla \phi, \Bnabla \Psi \right\rangle + Cn\left\langle \Bn \Bcdot \Bnabla \phi, \Psi \right\rangle_{\partial \Omega} =0 , \label{e:mu1} 
\end{align}
where $\langle \cdot, \cdot \rangle$ denotes the $L^{2}(\Omega)$  inner product.
The Navier-Stokes equation is solved by adopting Chorin’s method~\cite{Chorin_MC_1968} where we first ignore the pressure in  Eq.~(\ref{e:NS2}) which is then discretized using the finite difference method to compute the tentative velocity $\Bu_{T}$:
\begin{equation}
\frac{\Bu_{T}- \Bu^{n-1}}{\delta t_{n}} +  \Bu^{n-1} \Bcdot\Bnabla \Bu^{n-1}
=  \frac{1}{Re}\nabla^2 \Bu^{n-1} -\frac{1}{We} \phi \Bnabla \eta, \label{e:Ten_velocity}
\end{equation}
where $\delta t_{n}$ is the time step and $\Bu^{n}$ is the value of $\Bu$ at time $t_{n}$. This is corrected to obtain the final velocity $\Bu^{n}$ as
\begin{equation}
\frac{\Bu^{n} -\Bu_{T}}{\delta t_{n}} = - \Bnabla p^{n}. \label{e:velocity}
\end{equation}
The pressure $p^{n}$ at time $t^{n}$ is computed by taking the divergence of Eq.~(\ref{e:velocity}) and using the continuity equation: 
\begin{equation}
\frac{ \Bnabla\Bcdot\Bu_{T}}{\delta t_{n}} = \nabla^2 p^{n}. \label{e:pressure}
\end{equation}
Finally, the weak form of the Navier-Stokes equation (\ref{e:NS2}) is obtained by multiplying both Eq.~(\ref{e:Ten_velocity}) and (\ref{e:velocity}) by $\Xi$,  and Eq.~(\ref{e:pressure}) by $\Pi$. Integrating over the $\Omega$ we then compute the tentative velocity  $\Bu_{T}$, velocity $\Bu^{n}$, and pressure $p^{n}$ at time $t=t_{n}$:
\begin{align}
&\langle \frac{\Bu_{T}- \Bu^{n-1}}{\delta t_{n}},\Xi\rangle + \langle \Bu^{n-1} \Bcdot\Bnabla \Bu^{n-1} , \Xi \rangle
+ \frac{1}{Re}\langle  \Bnabla \Bu^{n-1}, \Bnabla \Xi \rangle = -\frac{1}{We}\langle \phi \Bnabla \eta , \Xi \rangle, \label{e:Ten_velocity2}\\
&\langle \Bu^{n},\Xi \rangle = \langle \Bu_{T}, \Xi \rangle -\delta t_{n}\langle \Bnabla p^{n}, \Xi \rangle,\label{e:pressure2}\\
&\langle \Bnabla p^{n}, \Bnabla \Pi \rangle = - \frac{1}{\delta t_{n}}\langle \Bnabla\Bcdot \Bu_{T},\Pi \rangle. \label{e:velocity2}
\end{align}

 %

\end{document}